\documentclass[11pt,a4paper]{article}
\pdfoutput=1
\usepackage{jheppub}
\notoc
\usepackage[T1]{fontenc}
\usepackage[spanish,USenglish]{babel}
\usepackage{bm}
\usepackage{microtype}
\usepackage{enumerate}
\usepackage{mathrsfs}
\usepackage{graphicx}

\begin{document}

\title{\texorpdfstring{$\chi_{c1}(2p)$}{chi c1 (2p)}: an overshadowed charmoniumlike resonance}
\author[a,c]{R. Bruschini}
\author[b,c]{P. Gonz\'alez}
\affiliation[a]{Department of Physics, The Ohio State University, Columbus, OH 43210, USA}
\affiliation[b]{\foreignlanguage{spanish}{Departamento de F\'isica Te\'orica, Universidad de Valencia}, 46100 Burjassot (Valencia), Spain} 
\affiliation[c]{\foreignlanguage{spanish}{Unidad Te\'orica, Instituto de F\'isica corpuscular} (UV--CSIC), 46980 Paterna (Valencia), Spain}
\emailAdd{bruschini.1@osu.edu}
\emailAdd{pedro.gonzalez@uv.es}
\arxivnumber{2207.02740}

\abstract{%
A thorough study of  the $J^{PC}=1^{++}$ elastic $D^{0}\bar{D}^{\ast0}$ and
$D^{+}D^{\ast-}$  scattering, where the form of the meson-meson interaction is inferred from
lattice QCD calculations of string breaking, is carried out for center-of-mass energies up to 4~GeV.
We show that the presence of $\chi_{c1}(3872)$, which can be naturally assigned
to either a bound or virtual charmoniumlike state close below
the $D^{0}\bar{D}^{\ast0}$  threshold, can overshadow a quasiconventional
charmoniumlike resonance lying above threshold.
This makes difficult the experimental detection of this resonance
through the $D^{0}\bar{D}^{\ast0}$ and
$D^{+}D^{\ast-}$  channels, despite being its expected main decay modes. We analyze
alternative strong and electromagnetic decay modes. Comparison with
existing data shows that this resonance may have already been observed through its decay  to  $\omega J/\psi$.
}

\maketitle

\section{Introduction}

Achieving a consistent and unified QCD-based explanation of quarkoniumlike mesons
\cite{PDG22} including both conventional ($Q\bar{Q}$ bound states where
$Q\equiv b$ or $c$) and unconventional states is a current challenge in hadronic
physics. In the last years, there has been a significant progress in this
respect. This progress is based, on the one hand, in effective field theories
and lattice analyses (see, for instance, \cite{Bra20} and references therein for a quite general review, and
\cite{Bra21}) and, on the other hand, in the development of quantum mechanical
formalisms where the potential incorporates inputs from lattice QCD \cite{Bic20,Bic21,Bru20, Bru21c}.
Henceforth we shall center on one of these formalisms, the so-called diabatic approach in QCD \cite{Bru20,Bru21c}.
In this approach, quarkoniumlike mesons with quantum numbers $J^{PC}$ are described through the solutions
of a multichannel Schr\"{o}dinger equation for $Q\bar{Q}$ and open-flavor meson-meson components,
where the form of the potential incorporates string breaking as observed in lattice QCD. This is a major
difference with quantum mechanical phenomenological analyses based on the same
degrees of freedom \cite{Eic80,Eic04,Eic06,Bar05,Bar08,Kal05,Pen07,Dan10b,Dan10a,Dong11,Fer13,Ort10,Ort18,Cin16,vanB21},
but where the $Q\bar{Q}$--meson-meson interaction lacks a
connection to QCD.

For energies below the lowest open-flavor meson-meson threshold, the solutions
of the Schr\"{o}dinger equation describe bound states which can be directly
assigned to quarkoniumlike resonances
stable against decays into open-flavor
meson-meson channels. Indeed, the spectrum obtained for
charmoniumlike and bottomoniumlike mesons far below the lowest threshold is
determined by the diagonal $Q\bar{Q}$ element of the diabatic potential matrix,
which is Cornell-like. Then, the incorporation of additional well-known spin
dependent corrections to the Cornell potential \cite{Eic81} allows for a good
description of existing data, see \cite{Bru21b} for a detailed discussion of
this point.

For energies above the lowest open-flavor meson-meson threshold, the solutions
of the Schr\"{o}dinger equation may contain one or more free-wave meson-meson
components. In such a case they have a natural interpretation in terms of
stationary open-flavor meson-meson scattering states. Then, scattering
resonances can be identified with quarkoniumlike mesons decaying into open-flavor
meson-meson channels. As a matter of fact, the diabatic analysis of the
scaled cross-section for elastic open-charm and open-bottom meson-meson
scattering with $J^{PC}=(0,1,2)^{++}$ and $1^{--}$
\cite{Bru21c} suggests that there is a spectrum of quasiconventional
resonances, in one-to-one correspondence with the spectrum of the
$Q\bar{Q}$ (Cornell) potential for energies above threshold, plus a
spectrum of unconventional states (scattering resonances and bound states), located close to some open-flavor meson-meson thresholds.

From the experimental point of view, although spectral states of both types
have been observed, some predicted quasiconventional resonances are missing.
In particular, the predicted quasiconventional $1^{++}$ charmoniumlike
resonance with a mass pretty close to that of the $2p$ $1^{++}$ Cornell
state, on which we shall center our analysis, is missing in the $c\bar{c}$ meson
listing of the Particle Data Group (PDG) \cite{PDG22}. From the
theoretical point of view, this particular prediction is strongly supported by
the following arguments. First, this resonance is located
within the region of applicability of the diabatic formalism (i.e., no
significant widths for nearby thresholds, no expected effects from
hybrids). Second, some spin-dependent terms in the diabatic
potential matrix, as for instance spin corrections to the Cornell potential
for $1^{++}$ states above the lowest threshold, are expected not to play a
significant quantitative role (for $(0,2)^{++}$ these effects are expected to be quantitatively more important \cite{Eic81}).
Third, this resonance comes out from the same
interaction giving rise to an unconventional state just below the
$D\bar{D}^{\ast}$ threshold which is naturally assigned to the well-established $\chi_{c1}(3872)$.

Notwithstanding these arguments, the fact that in \cite{Bru21c} the mass
difference between the $D^{0}\bar{D}^{\ast0}$and $D^{+}D^{\ast-}$ thresholds was neglected,
and the big difference between the effective value of the $c\bar{c}$--meson-meson interaction strength and the energy gap calculated in lattice QCD,
could cast some doubts about the robustness of this prediction.

In order to overcome these shortcomings, in this article we implement the
$D^{0}\bar{D}^{\ast0}$ and $D^{+}D^{\ast-}$ threshold
mass difference and we make a simplified analysis of possible alternative
dynamical mechanisms giving rise to $\chi_{c1}(3872)$. In this manner, we
confirm the prediction of a quasiconventional resonance in the elastic $D^{0}\bar{D}^{0\ast
}$ and $D^{+}D^{\ast-}$ scattering channels, that we call $\chi
_{c1}(2p)$. As a main novelty from this article, we show that the effect of this resonance on the scattering
cross-section may be overshadowed by that of $\chi_{c1}(3872)$, making difficult
its experimental detection through the $D^{0}\bar{D}^{\ast0}$ and
$D^{+}D^{\ast-}$ channels despite being its expected dominant decay modes.
This may explain why $\chi_{c1}(2p)$ is missing in the PDG listing and
suggests that alternative detections, such as from OZI-forbidden
decays, could be instrumental for its experimental establishment.

It should be pointed out that the prediction of $\chi_{c1}(3872)$ and
$\chi_{c1}(2p)$ states as mainly due to the meson-meson channel coupling to
the charmonium (charm-anticharm) $2p$ component has being previously reported
by other authors using different calculation frameworks, see, for instance,
\cite{Kal05,Dan10b,Ort10} for a panoramic of the different results obtained. In all these references, a ${^{3}\!P}_0$-type string breaking
mixing with no connection to unquenched lattice QCD results is used. In \cite{Kal05},
the coupling generates the $\chi_{c1}(2p)$ resonance (with a mass about $3990$
MeV) with a relatively narrow width, and a virtual state pretty close to
threshold assigned to $\chi_{c1}(3872)$. In \cite{Ort10}, the coupling gives rise to a
bound state close to threshold assigned to $\chi_{c1}(3872)$ and a $\chi
_{c1}(2p)$ resonance (with a mass about $3936$ MeV) tentatively assigned to
$X(3940)$. In this case, no width for $\chi_{c1}(2p)$ has been reported.
Finally, in \cite{Dan10b} a pole analysis of the coupled-channel system and a
calculation of the $D\bar{D}^{\ast}$ production cross-section is carried
out. This calculation shows that the coupling between $\chi_{c1}(2p)$ and the 
$D^{0}\bar{D}^{\ast 0}$ threshold may transform the original charmonium state into a very broad resonance above
threshold while generating a steep rise near threshold, associated to $\chi_{c1}(3872)$.

As mentioned before, the main difference between our study and previous
analyses is the use of a diabatic framework which allows for the
implementation of a mixing interaction whose form is derived from (unquenched)
lattice calculations of string breaking. As we shall show, a bound state just
below the $D^{0}\bar{D}^{\ast 0}$ threshold, which is assigned to
$\chi_{c1}(3872)$, and a $\chi_{c1}(2p)$ resonance with a mass about $3950$ MeV
and a width of around $70$ MeV come out. However,
the $\chi_{c1}(2p)$ is hardly visible in
the scattering cross-section. This lack of
visibility is essentially different from that in the production cross-section calculated in \cite{Dan10b}.
In fact, we find that the $\chi_{c1}(2p)$ may be relatively narrow, but effectively overshadowed in the
open-charm channels by the tail of the threshold enhancement due to $\chi_{c1}(3872)$.
This, unlike a very broad $\chi_{c1}(2p)$, could explain its lack of detection in
open-charm channels while leaving the door open to its possible detection through alternative channels, as commented above.

The contents of this article are organized as follows. In section~\ref{diabsec} we proceed to a brief
review of the main physical ingredients of the diabatic approach in QCD and
focus on its calculation of an unconventional state just below the
$D^{0}\bar{D}^{\ast0}$ threshold. In section~\ref{scatsec} we review the main aspects
of the application of the diabatic approach to open-charm meson-meson
scattering. In section~\ref{csecsec} we calculate the $J^{PC}=1^{++}$ elastic
$D^{0}\bar{D}^{\ast0}$ and $D^{+}D^{\ast-}$ scattering cross-sections for center-of-mass energy up to 4~GeV
under the assumption that the main physical mechanism generating
$\chi_{c1}(3872)$ is string breaking. We proceed to a physical
analysis of the structures in the cross-sections and to a study of their
robustness against variations of the strength of the string breaking interaction
potential responsible for the scattering. The main outcome is the prediction
of an overshadowed quasiconventional scattering resonance at about 3960~MeV.
In section~\ref{toysec}, we examine possible alternative scenarios to generate the $\chi_{c1}(3872)$, by adding different physical mechanisms to a diabatic potential with a smaller strength of the string breaking interaction. On the one hand, we consider the addition of a pion exchange meson-meson potential. On the other hand, we analyze the addition of a new compact channel. The overshadowing of  $\chi_{c1}(2p)$ is realized in the pion exchange scenario as well.  In contrast, no theoretical evidence of overshadowing is found in the compact case. In section~\ref{chisec}, we focus on the overshadowing scenario that could explain the current lack of experimental candidates for $\chi_{c1}(2p)$. We examine favored hidden-charm decay modes of this resonance and discuss about the possibility of missing $J^{PC}=(0,2)^{++}$ partners in the same energy region.
Finally, in section~\ref{sumsec} we summarize our main achievements.

\section{Diabatic approach in QCD}
\label{diabsec}

In the diabatic approach in QCD \cite{Bru20,Bru21c}, a $J^{PC}$
quarkoniumlike meson is made of $Q\bar{Q}$ and open-flavor meson-meson
components ($M_{1}^{(i)}\bar{M}_{2}^{(i)}$ with $i=1,2,\dots$)
obeying a coupled-channel Schr\"{o}dinger equation. The form of
the diabatic potential matrix entering in this equation is obtained from
lattice calculations of the static energy levels for $Q\bar{Q}$ and for
mixed $Q\bar{Q}$-$M_{1}^{(i)}\bar{M}_{2}^{(i)}$ configurations.

More precisely, the dynamics is described by the diabatic potential matrix
\begin{equation}
\mathrm{V}(\bm{r}) =
\begin{pmatrix}
V_{Q\bar{Q}}(r)				& V_{\textup{mix}}^{(1)}(\bm{r})	& V_{\textup{mix}}^{(2)}(\bm{r})	& \hdots 	& V_{\textup{mix}}^{(N)}(\bm{r})	\\
V_{\textup{mix}}^{(1)}(\bm{r})^\dag	& T^{(1)}					&						&  		& 						\\
V_{\textup{mix}}^{(2)}(\bm{r})^\dag	& 						& T^{(2)}					& 	 	& 						\\
\vdots						& 						& 						& \ddots 	&  						\\
V_{\textup{mix}}^{(N)}(\bm{r})^\dag	& 						& 						& 	 	& T^{(N)}					\\
\end{pmatrix}
\label{diabpot}
\end{equation}
where$V_{Q\bar{Q}}(r)$ and $T^{(i)}$ are, respectively, the diagonal elements corresponding to the $Q\bar{Q}$ and $M_{1}^{(i)}\bar{M}_{2}^{(i)}$ components, $V_{\textup{mix}}^{(i)}(\bm{r})$ the mixing potential between them, and vanishing matrix elements have been omitted for simplicity. Notice that we have conveniently assumed that the diagonal meson-meson matrix elements are simply given by the threshold masses,
\[
T^{(i)}=m_{M_{1}^{(i)}}+m_{\bar{M}_{2}^{(i)}}
\]
where $m_{M_{1}^{(i)}}$ and $m_{\bar{M}_{2}^{(i)}}$ are the masses of the
corresponding mesons, and that there are no offdiagonal potential matrix elements mixing different meson-meson components with each other (we shall come back to this point in section~\ref{toysec}). In what follows, we shall use the experimental open-charm
meson masses quoted in \cite{PDG22}.

The so-called adiabatic-to-diabatic transformation, as explained in more detail in \cite{Bru20} and references therein, allows to draw a direct correspondence between the diabatic potential matrix \eqref{diabpot} and the static energy levels of static $Q$ and $\bar{Q}$ sources calculated in lattice QCD. Namely, the static energy levels are just the eigenvalues of \eqref{diabpot}. Then, one may use quenched (including only $Q\bar{Q}$) and unquenched (including $Q\bar{Q}$ and meson-meson configurations) lattice energy levels for determining the diabatic potential matrix elements via the following procedure:
\begin{enumerate}
\item\label{step1} Quenched lattice results \cite{Bal01} are incorporated through
the diagonal $Q\bar{Q}$ diabatic potential matrix element, $V_{Q\bar{Q}}(r)$.
\item\label{step2} Unquenched lattice results \cite{Bal05,Bul19} are combined with $V_{Q\bar{Q}}(r)$ and the assumption on the meson-meson potentials to determine the offdiagonal $Q\bar
{Q}$-$M_{1}^{(i)}\bar{M}_{2}^{(i)}$ diabatic potential matrix elements.
\end{enumerate}

From step number \ref{step1}, one can write write $V_{Q\bar{Q}}(r)$ as a
confining Cornell-like potential
\[
V_{Q\bar{Q}}(r) = \sigma r-\frac{\chi}{r}+m_{Q}+m_{\bar{Q}}-\beta
\]
with the flavor-independent parameters $\sigma$ and $\chi$, respectively,
the string tension and color Coulomb strength.
Notice that the arbitrary constant
in the potential has been conveniently written in terms of the heavy quark and
antiquark masses, $m_{Q}$ and $m_{\bar{Q}}$, plus a constant $\beta$ which
can also be taken as flavor-independent through a suitable
choice of the masses.

For the case $Q=c$ we shall deal with, standard phenomenological values of the
parameters are \cite{Eic94}
\begin{align*}
\sigma &  =925.6\text{~MeV/fm},\\
\chi &  =102.6\text{~MeV~fm},\\
m_{c}  &  =1840\text{~MeV},\\
\beta &  =855\text{~MeV},
\end{align*}
allowing for a quite accurate spectral description of (conventional)
charmonium once spin-dependent corrections are implemented \cite{Eic81}. It
should be also remarked that the value of $m_{c}$ is very much constrained by
the requirement of describing properly electromagnetic transitions
\cite{Bru19b}.

As for the offdiagonal $Q\bar
{Q}$-$M_{1}^{(i)}\bar{M}_{2}^{(i)}$ diabatic potential matrix elements, or
mixing potentials, from step number \ref{step2} their radial part%
\footnote{The spin-angular part of the mixing potential has a complicated analytical expression, which also depends on the representation of the cylindrical symmetry group of Born-Oppenheimer. Its determination, way beyond the scope of the current study, shall be detailed in a future paper.}
can be approximated by the analytical form \cite{Bru20,Bru21c}
\begin{equation}
\label{mixpot}
V_{\textup{mix}}^{(i)}(r) = - \frac{\Delta_{Q}}{2}\exp\biggl\{-\frac
{(V_\textup{C}(r)-T^{(i)})^{2}}{2\sigma^{2}\rho^{2}}\biggr\}
\end{equation}
where $\Delta_{Q}$ and $\rho$ are, respectively, an effective mixing strength and radial scale parameter, assumed to be equal
equal for all thresholds.

It is worth emphasizing that this form of the mixing potential incorporates the fact that
the $Q\bar{Q}$-$M_{1}^{(i)}\bar{M}_{2}^{(i)}$ mixing is only
significant in an interval around the crossing radius $r_\textup{c}^{(i)}$, defined by $V_\textup{C}(r_\textup{c}^{(i)})=T^{(i)}$,
the size of this interval being determined by the value of $\rho$,
which has been fixed to
\[
\rho_\textup{c}=0.3\text{~fm}
\]
in order to reproduce the behavior of the mixing angle between $Q\bar{Q}$
and meson-meson observed in lattice QCD \cite{Bru20}.

Regarding the value of $\Delta_{c}$, it is fixed from the requirement that the solution of the diabatic Schr\"odinger equation with this diabatic potential matrix gives rise to a $1^{++}$ bound state (this is, a charmoniumlike meson stable against decays into an open-charm di-meson pair) just below the $D^{0}\bar{D}^{\ast0}$
threshold, in correspondence with the well-established experimental state
$\chi_{c1}(3872)$. This gives
\begin{equation}
\Delta_{c}=102.2\text{~MeV}.
\label{DeltacDif}
\end{equation}

\begin{figure}
\centering
\includegraphics{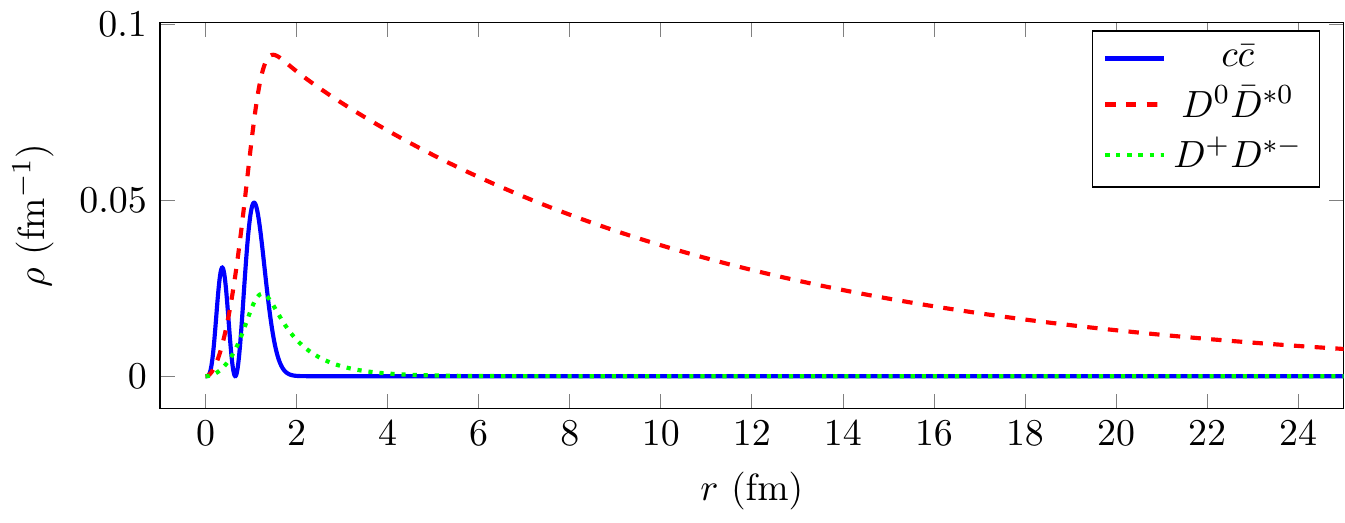}
\caption{\label{wfunc}Radial probability densities for the bound state at 3871.6~MeV. Meson-meson components with higher threshold are negligible. The $D^0 \bar{D}^{\ast 0}$ component extends far beyond the right limit of the plot, reaching up to $r \approx 70$ fm.}
\end{figure}

Some comments are in order. First, in some previous applications of the
diabatic approach to charmoniumlike mesons, see for instance
\cite{Bru20,Bru21c}, the mass difference between $D^{0}\bar{D}^{\ast0}%
$and $D^{+}D^{\ast-}$ was neglected, so that the effect of both
thresholds was taken into account through one effective isospin-zero threshold
with the measured mass of $D^{0}\bar{D}^{\ast0}$. As a consequence, the
effective mixing strength (also called $\Delta_{c}$) had a quite larger value.
Second, the calculated bound state with a mass of
$3871.6$ MeV, just below the $D^{0}\bar{D}^{\ast0}$ threshold, and with a
$93\%$, $4\%$, and $3\%$ probabilities for the $D^{0}\bar{D}^{\ast0}$,
$c\bar{c}$, and $D^{+}D^{\ast-}$ components, respectively, can
be naturally assigned to the well-established $\chi_{c1}(3872)$ since this
composition is in line with its observed decay properties. Hence, $\chi
_{c1}(3872)$ can be approximately seen as a loosely bound $D^{0}\bar{D}^{\ast0}$ state.
For the sake of completeness, we plot in figure~\ref{wfunc} the radial probability densities
for $c\bar{c}$, $D^{0}\bar{D}^{\ast0}$, and $D^{+}D^{\ast-}$.

It is also worth mentioning that within the diabatic framework the value of
$\Delta_{Q}$ can be related to $\sigma$ and $m_{Q}$ through a semiempirical
scaling law \cite{Bru22}. We could have used this scaling law to get the same value of
$\Delta_{c}$ from the phenomenological values of $\Delta_{b},$ $m_{b}$ and
$m_{c}$, so that the calculated bound state assigned to $\chi_{c1}(3872)$
would have been a prediction. However, as this scaling law makes unclear the
connection with lattice QCD, see section~\ref{toysec}, in this study we have preferred to fix $\Delta_{c}$
in a purely phenomenological manner.

\section{Open-flavor meson-meson scattering}
\label{scatsec}

For energies above the lowest $J^{PC}$ meson-meson threshold, $T^{(1)}$, the solutions of
the Schr\"{o}dinger equation contain one or more free-wave meson-meson
components, $M_{1}^{(j)}\bar{M}_{2}^{(j)}$. Since their wave function is not normalizable in the
traditional sense, the interpretation of these states as quarkoniumlike mesons is somewhat more intricate.
As a matter of fact, one has a continuum spectrum of solutions with energies $E > T^{(1)}$, instead of a discrete
spectrum of quarkoniumlike states with their masses. These continuum solutions have a natural interpretation
in terms of stationary $J^{PC}$ meson-meson scattering states where the
scattering process is induced by the mixing potential: $M_{1}^{(j^{\prime}%
)}\bar{M}_{2}^{(j^{\prime})}\rightarrow(Q\bar{Q})
\rightarrow M_{1}^{(j)}\bar{M}_{2}^{(j)}$. Then, scattering resonances
can be used to identify quarkoniumlike mesons decaying into open-flavor
meson-meson channels.

More concretely, the asymptotic behavior of the $J^{PC}$
solutions with energy $E$, $\psi_{J, m_{J}; h}^{(j)}(\bm{r})$ with $m_J$ the projection of $\bm{J}$ and $j$ and $h$ labeling the meson-meson components and linearly independent solutions with the same energy, respectively, can be expressed in terms of the asymptotic behavior of $J^{PC}$
stationary meson-meson scattering states with the same energy, $\psi_{J^{PC}, m_{J}; k^{\prime}}^{j \leftarrow j^{\prime}}(\bm{r})$ with $j$ and $(j^\prime, k^\prime)$ labeling, respectively, the ingoing channel and outgoing partial-wave channel, and vice versa,
\[
\psi_{J^{PC}, m_{J}; k^{\prime}}^{j \leftarrow j^{\prime}}(\bm{r}) = \sum_{h} \psi_{J, m_{J}; h}^{(j)}(\bm{r}) \Gamma_{J^{PC}; k^{\prime}; h}^{(j^{\prime})}
\]
where $\Gamma_{J^{PC}; k^{\prime}; h}^{(j^{\prime})}$ are change of basis matrix elements.
From this expression, one can show that the $J^{PC}$ meson-meson $S$-matrix is given in terms of the Jost matrices $\mathscr{F}_{J^{PC}}^{\pm}$ as
\[
S_{J^{PC}} = \mathscr{F}^{+}_{J^{PC}} \bigl(\mathscr{F}^{-}_{J^{PC}}\bigr)^{-1}.
\]

As the Jost matrices are completely determined from the numerical solutions of the coupled-channel Schr\"odinger equation above threshold,$\psi_{J, m_{J}; h}^{(j)}(\bm{r})$, one can thus obtain the numerical values of the elastic ($j=j^{\prime}$) and inelastic ($j\neq
j^{\prime}$) $J^{PC}$ meson-meson scattering amplitudes without any additional parameter. From
them, the corresponding $J^{PC}$ cross-sections, $\sigma_{J^{PC}}^{j\leftarrow
j^{\prime}}$ are calculated and the resonant structures identified.

We refer the interested reader to \cite{Bru21c} for the technical details of the procedure outlined above, as well as the numerical procedure to calculate the solutions above threshold.

It is worth to mention that the identification of some structures from cross-section data may be easier, as will be shown below,
through the use of the scaled cross-section defined as
\begin{align*}
\bar{\sigma}_{J^{PC}}^{j\leftarrow j^{\prime}}  &  =\frac
{(2s_{M_{1}^{(j^{\prime})}}+1)(2s_{M_{2}^{(j^{\prime})}}+1)}{4\pi
(2J+1)}(p^{(j)})^{2}\sigma_{J^{PC}}^{j\leftarrow j^{\prime}}\\
&  =\sum_{k,k^{\prime}}\bigl\lvert p^{(j)}f_{J^{PC};k,k^{\prime}}^{j\leftarrow
j^{\prime}}\bigr\rvert ^{2}%
\end{align*}
and satisfying
\[
\sum_{j,j^{\prime}}\bar{\sigma}_{J^{PC}}^{j\leftarrow j^{\prime}}\leq1
\]
where $s$ stands for the spin, $p^{(j)}$ for the modulus of the relative
momentum between $M_{1}^{(j)}$ and $\bar{M}_{2}^{(j)}$, $k$ for the
partial wave $(l^{(j)},s^{(j)})$ coupling to $J^{PC}$, and $f$
for the scattering amplitude.

Henceforth, we focus on the $J^{PC}=1^{++}$ open-charm meson-meson scattering
for center-of-mass energy up to 4~GeV. This energy limit makes us confident
that charmonium hybrids, expected to be higher in energy, and higher
thresholds, even with a large width as for example $D\bar{D}_{1}$, do not
play any quantitative role.

In order to take into account the distinction between the $D^{0}\bar{D}^{\ast0}$
and $D^{+}D^{\ast-}$ thresholds, we consider the two distinct $D^{0}%
\bar{D}^{\ast0}$ and $D^{+}D^{\ast-}$ channels, instead of
one isospin-zero $D\bar{D}^{\ast}$ channel with an effective
threshold as done in \cite{Bru21c}.

\section{\texorpdfstring{$1^{++}$}{1++} scaled cross-sections}
\label{csecsec}

\begin{figure}
\centering
\includegraphics{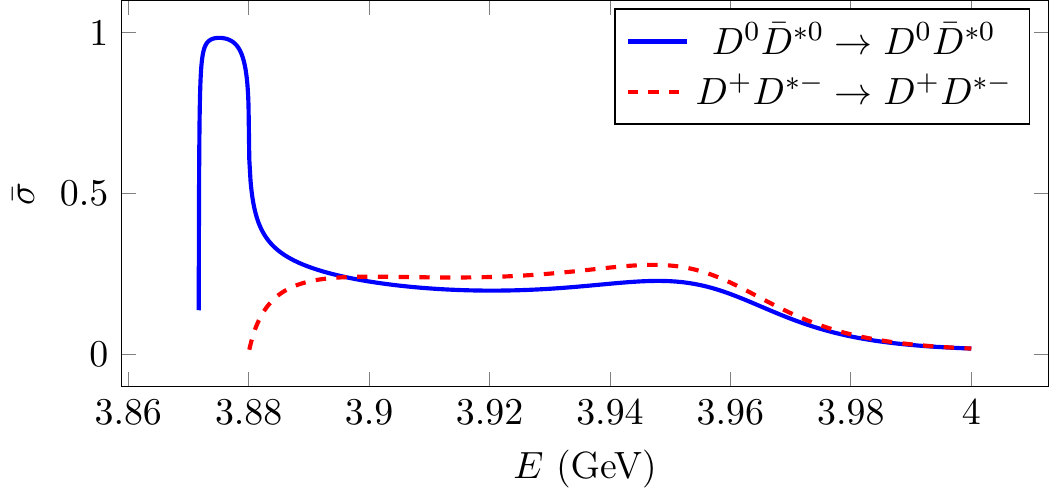}
\caption{\label{scaledcsec102}Elastic scaled cross-sections for $J^{PC}=1^{++}$, calculated with $\Delta_c=102.2$~MeV.}
\end{figure}

The calculated scaled cross-sections for $J^{PC}=1^{++}$ $D^{0}\bar
{D}^{\ast0}$ and $D^{+}D^{\ast-}$ elastic scattering up to 4~GeV
center-of-mass energy are drawn in figure~\ref{scaledcsec102}. Let us point out that the
closed $D^{\ast0}\bar{D}^{\ast0}$ and $D^{+\ast}D^{\ast-}$ channels,
with thresholds slightly above 4~GeV, have been also taken into account in the multichannel Schr\"{o}dinger equation.

The figure shows clearly some structures. Next, we proceed to a detailed
physical analysis of them through the use of Argand diagrams. Furthermore, we
analyze the robustness of these structures under variations of the
value of the mixing strength parameter.

\subsection{\texorpdfstring{$D^{0}\bar{D}^{\ast0}\rightarrow D^{0}\bar{D}^{\ast0}$}{D0D0starD0D0star}
scattering}

A look at figure~\ref{scaledcsec102} shows that the $D^{0}\bar{D}^{\ast0}$ elastic scattering
scaled cross-section presents two enhacements, a first big one just above
threshold and a second smaller one at a higher energy.

\begin{figure}
\centering
\includegraphics{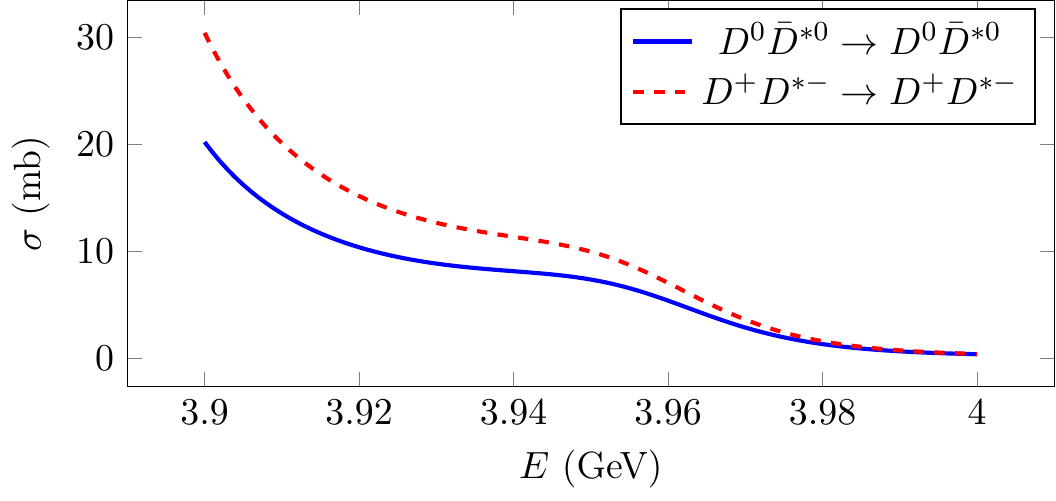}
\caption{\label{natcsec102}Elastic cross-sections for $J^{PC}=1^{++}$ near 3960~MeV, calculated with $\Delta_c=102.2$~MeV.}
\end{figure}

It is worth to mention that if we plotted the (non-scaled) cross-section
$\sigma_{1^{++}}^{D^{0}\bar{D}^{\ast0}\leftarrow D^{0}\bar{D}%
^{\ast0}}$ instead of $\bar{\sigma}_{1^{++}}^{D^{0}\bar
{D}^{\ast0}\leftarrow D^{0}\bar{D}^{\ast0}}$, then the second enhancement
would be hardly visible as a shoulder of the order of
millibarns, see figure~\ref{natcsec102},
to be compared with the threshold enhancement of the order of
hundreds of barns.

In order to physically analyze these enhancements, let us realize first that
just above threshold the scattering is dominated by the $s$-wave,
$l_{D^{0}\bar{D}^{\ast0}}=0$, so that 
\[
\bar{\sigma}_{1^{++}}^{D^{0}\bar{D}^{\ast0}\leftarrow D^{0}%
\bar{D}^{\ast0}}=\lvert pf_{0}\rvert ^{2}%
\]
where $p$ is the relative $D^{0}\bar{D}^{\ast 0}$ momentum and $f_0$ the $s$-wave elastic $D^{0}\bar{D}^{\ast 0}$ scattering amplitude.

Let us recall that for $\Delta
_{c}=102.2$~MeV there is a bound state solution at 3871.6~MeV,
close below the $D^{0}\bar{D}^{\ast0}$ threshold $T^{(D^{0}\bar
{D}^{\ast0})}=3871.7$~MeV. This bound state has been assigned to $\chi_{c1}(3872)$.
As it is well known from scattering theory, the presence of the shallow
bound state determines the behavior of the amplitude at low momenta (see, for example, \cite{Wei15}). More
precisely, if we define the binding momentum
\[
\alpha= \sqrt{2\mu_{D^{0}\bar{D}^{\ast0}}(T^{(D^{0}%
\bar{D}^{\ast0})}-E)}
\]
where $\mu_{D^{0}\bar{D}^{\ast0}}$ is the $D^{0}\bar{D}^{\ast0}$
reduced mass and $E$ is the total energy of the bound state at rest (i.e., its mass), then for $p\ll\alpha$ one has
\begin{equation}
f_{0} \simeq \frac{i}{p-i\alpha}\simeq-\frac{1}{\alpha}+i\frac{p}{\alpha^{2}},
\label{LM}
\end{equation}
so that
\[
\Re[pf_{0}] \leq 0
\]
and%
\[
\bar{\sigma}_{1^{++}}^{D^{0}\bar{D}^{\ast0}\leftarrow
D^{0}\bar{D}^{\ast0}} \simeq \frac{p^{2}}%
{p^{2}+\alpha^{2}}\simeq\frac{p^{2}}{\alpha^{2}}.
\]
From these expressions it is clear that the bound state implies a pole of the
scattering amplitude at positive imaginary momentum $i\alpha$.%
\footnote{In contrast, if one had an $s$-wave 
virtual state with energy $E$ on the unphysical Riemann sheet, then the scattering
amplitude would have a pole at negative imaginary momentum
$-i\alpha$. Then, $f_{0} \approx \frac{1}{\alpha} + i\frac{p}{\alpha^{2}}$ and $\Re[pf_{0}] \geq 0$ for $p \ll \alpha$.
In case of a zero-binding-energy state at threshold, one would have
$\alpha=0$ and $p f_0$ would be pure imaginary ($\Re[p f_0]=0$) at low momentum.
Notice that the cross-section at low momentum has the same expression
for the bound and virtual state cases, whereas it diverges as
$p^{-2}$ in the case of a zero-binding-energy state.}

\begin{figure}
\centering
\includegraphics{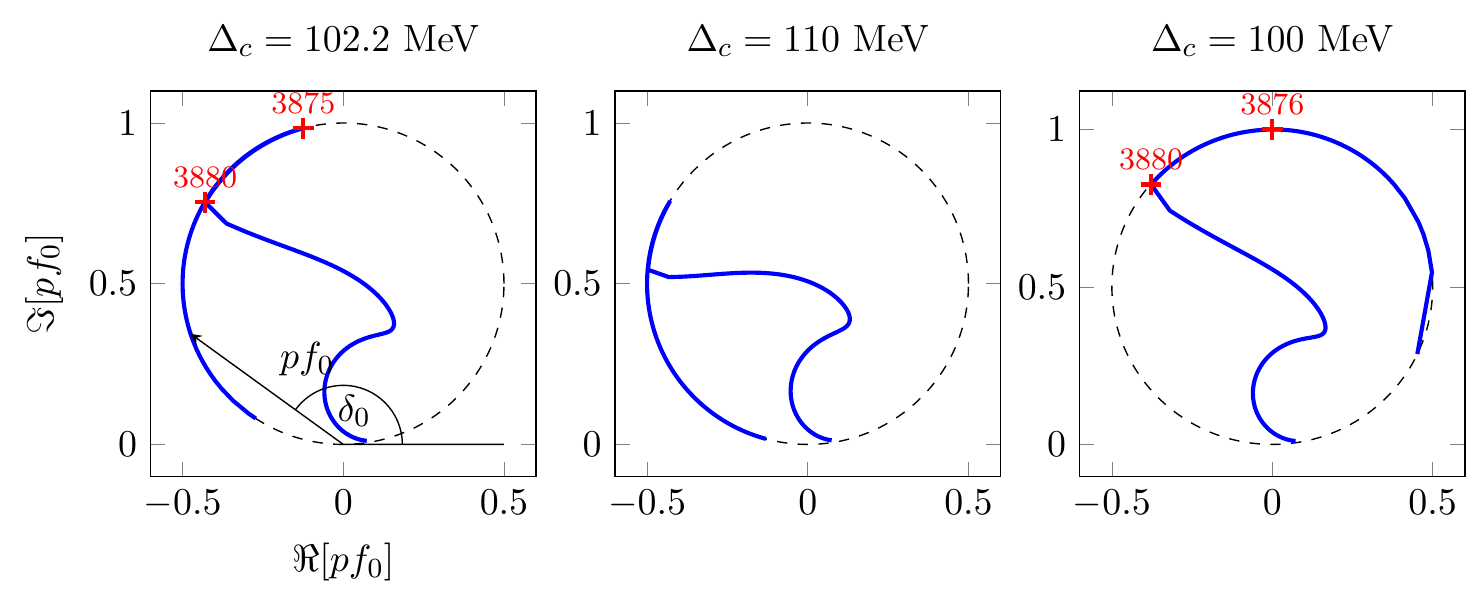}
\caption{\label{argandD0D0asts}Argand diagram for the $s$-wave elastic $D^0\bar{D}^{\ast0}$ scattering amplitude, calculated with different values of $\Delta_c$. Some energies in MeV, referred to in the text, are highlighted.}
\end{figure}

The analytical expression of the scattering amplitude for $p \ll \alpha$, Equation~\eqref{LM}, can be compared to what we calculate numerically from the diabatic Schr\"odinger equation. Indeed, this low-momentum behavior is clearly illustrated by the $s$-wave Argand diagram for the scaled scattering amplitude $pf_{0}$, left panel in figure~\ref{argandD0D0asts}.

The diagram follows clockwise the unitary circle from threshold to a center of
mass energy of about 3875~MeV,
indicating the complete dominance of the $s$-wave up to this energy. (Notice
that the clockwise behavior at low momentum is evident from the expression of
the amplitude \eqref{LM}.) Then, for low momenta we can use the effective
range expansion%
\begin{equation}
p\cot\delta_{0}=-\frac{1}{a_{0}}+\frac{r_{0}}{2}p^{2}+\dots \label{ER}%
\end{equation}
where $\delta_{0}$ is the $s$-wave phase shift, $a_{0}$ the scattering length, and $r_{0}$ the so-called
effective range.

Notice that, using the well-known relation
\begin{equation}
p\cot\delta_{0}=\frac{1}{f_{0}}+ip \label{REL}%
\end{equation}
and the former expression of the amplitude at low momentum \eqref{LM}, one has
\[
\lim_{p\rightarrow0} p\cot\delta_{0} = -\alpha
\]
and therefore $a_0 = \alpha^{-1}$. Indeed, from the numerical values of $f_{0}$ and the corresponding values of
$p\cot\delta_{0}$ from \eqref{REL}, the fitting of \eqref{ER} at low momentum
gives
\[
a_{0}  = 18.8~\text{fm}
\]
reflecting the presence of the shallow bound state at 3871.6~MeV. As for the effective range, one obtains
\[
r_{0}  = 0.3~\text{fm},
\]
which is of the order of the range of the meson-meson force mediated by the mixing
potential.

Beyond the low-momentum region $p \ll \alpha$, Equation~\eqref{LM} ceases to be valid, and the physical information cannot be summarized by $a_0$ and $r_0$ alone. Instead, the physical picture must be reconstructed from the complicated numerical behavior of the scattering amplitude.

When increasing $p$, the modulus of $pf_{0}$ (or, equivalently, the scaled cross-section) increases its value reaching a maximum at an energy of about 3875~MeV, corresponding to the peak of the big enhancement in the scaled cross-section.

\begin{figure}
\centering
\includegraphics{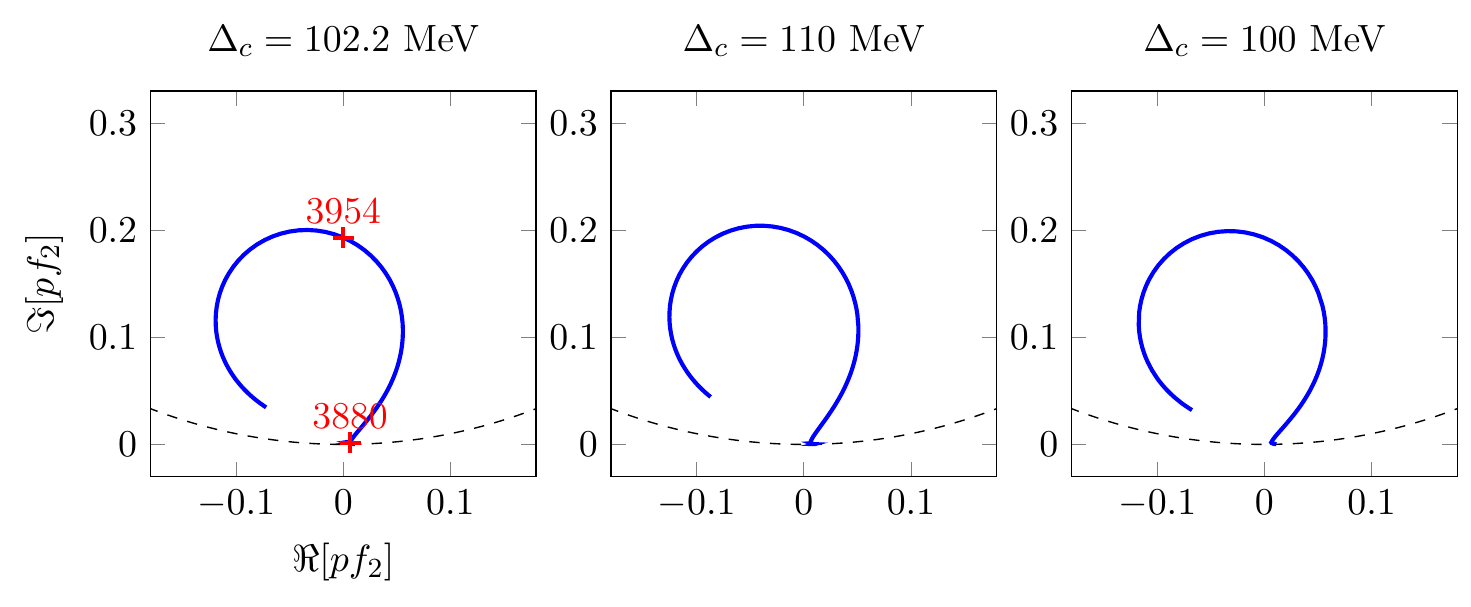}
\caption{\label{argandD0D0astd}Argand diagram for the $d$-wave elastic $D^0\bar{D}^{\ast0}$ scattering amplitude, calculated with different values of $\Delta_c$. Some energies in MeV, referred to in the text, are highlighted.}
\end{figure}

When going through 3875~MeV, the $s$-wave Argand diagram changes
from clockwise to counterclockwise behavior still following the unitary circle
up to an energy about 3880~MeV, at which the inelastic channel $D^{0}\bar{D}^{\ast0} \to D^{+}D^{\ast-}$ opens. Then $pf_{0}$
stops following the unitary circle. At approximately this same energy the $d$-
wave, $l_{D^{0}\bar{D}^{\ast0}}=2$, starts to give a non-negligible
contribution. This can be seen more clearly in the Argand diagram
for the $d$-wave elastic $D^{0}\bar{D}^{\ast0}$ scattering amplitude $f_2$, left panel in figure~\ref{argandD0D0astd}.

A look at this diagram shows that the $d$-wave phase shift $\delta_{2}$ experiences a
counterclockwise variation from $0$ to $\pi$ passing $\frac{\pi}{2}$ at an
energy about 3954~MeV, signaling the presence of a standard Breit-Wigner resonance centered at
this energy. This resonance, which can be also inferred from the $s$-wave
Argand diagram although not in such a clear way due to the presence of a
background phase shift, gives rise, in combination with the tail of the big enhancement, to the smaller enhancement observed in the scaled cross-section
$\bar{\sigma}_{1^{++}}^{D^{0}\bar{D}^{\ast0}\leftarrow D^{0}\bar{D}^{\ast0}}$. We call this resonance $\chi_{c1}(2p)$, since it corresponds to a quasiconventional $c\bar{c}$ state coming out (as it was also the case for
$\chi_{c1}(3872)$) from the interaction of the conventional $2p$ charmonium state
from the Cornell potential, which has a mass of
3953~MeV, with the $D^{0}\bar{D}^{\ast0}$ and $D^{+}D^{\ast-}$ thresholds. From
the calculated scaled cross section, the width of the $\chi_{c1}(2p)$ can be estimated to be $70\pm10$~MeV.

\subsection{Parameter dependence}
\label{pardep}

It is illustrative to analyze how the above results change under variations of the effective mixing strength
$\Delta_{c}$.

Let us first realize that by increasing $\Delta_{c}$ one just increases the binding energy of the state. So, increasing $\Delta_{c}$ from 102.2~MeV to 110~MeV, the mass of the bound state goes from
3871.6~MeV to 3870.4~MeV.

\begin{figure}
\centering
\includegraphics{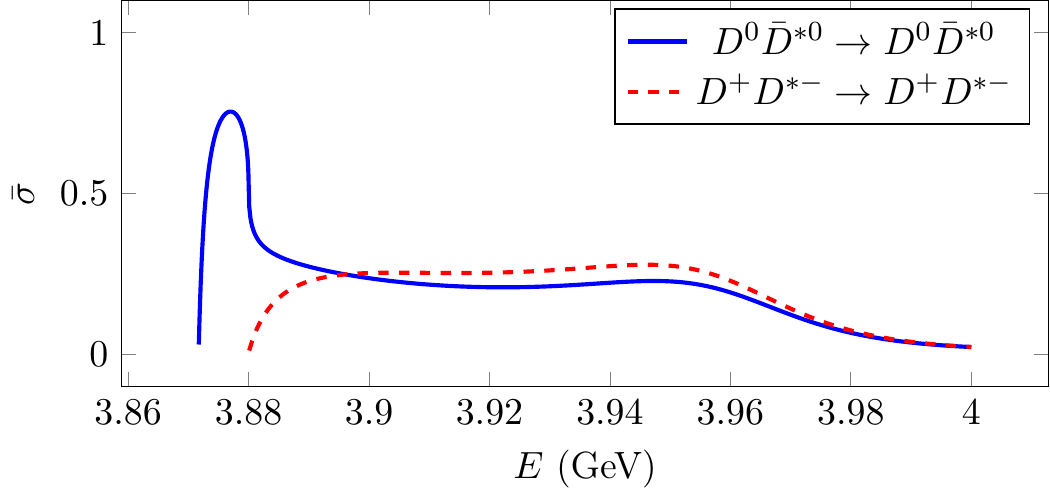}
\caption{\label{scaledcsec110}Elastic scaled cross-sections for $J^{PC}=1^{++}$, calculated with $\Delta_c=110$~MeV.}
\end{figure}

As can be checked, the Argand
diagrams drawn in the central panels of Figs.~\ref{argandD0D0asts} and \ref{argandD0D0astd} and the scaled cross-section plotted in figure~\ref{scaledcsec110} are qualitatively similar to those in the left panels of Figs.~\ref{argandD0D0asts} and \ref{argandD0D0astd}, and in figure~\ref{scaledcsec102}, respectively.

The modest quantitative differences have to do with a slight decrease of the
scattering length giving rise to a smaller value of the scaled cross-section
at low energy.

On the other hand, the binding energy gets closer to zero when decreasing $\Delta_{c}$ from 102.2~MeV to 101~MeV.

If we continue decreasing $\Delta_{c}$, at some value close below 101~MeV
the binding energy vanishes and the Argand diagram changes dramatically at
$p=0$. In that case, one has $pf_{0}=i$ (i.e., $\delta_{0}=\frac{\pi}{2}$) at $p=0$, meaning that the amplitude
has a pole there. This situation, which cannot be realized exactly in numerical calculations, corresponds to a zero-binding-energy state at threshold
giving rise to an infinite scattering length.

As for lower values of the mixing strength, such as $\Delta_{c}=100$~MeV, the
low-momentum scattering amplitude shows a counterclockwise behavior with
$\Re[pf_{0}]>0$ as it correponds to a virtual
state, see right panel in figure~\ref{argandD0D0asts}.

\begin{figure}
\centering
\includegraphics{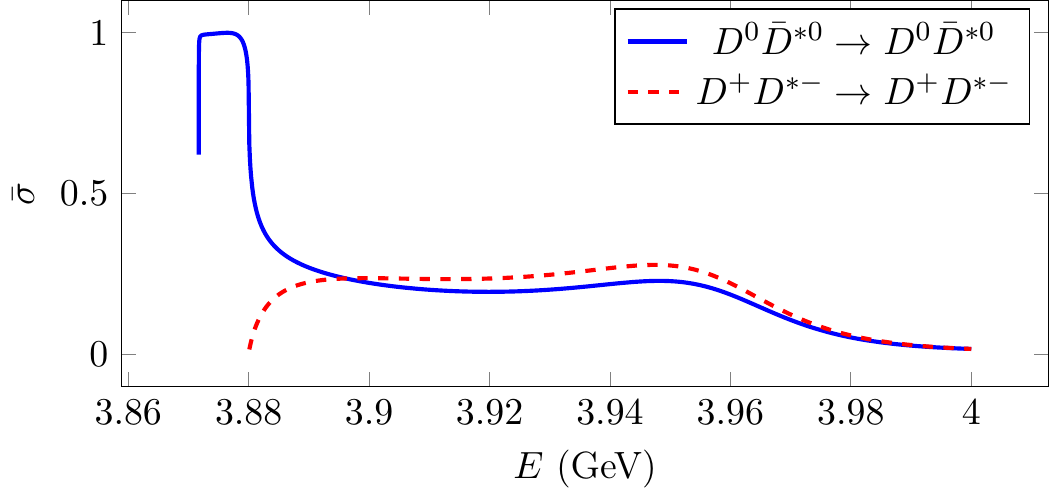}
\caption{\label{scaledcsec100}Elastic scaled cross-sections for $J^{PC}=1^{++}$, calculated with $\Delta_c=100$~MeV.}
\end{figure}

The physical interpretation is that although the attractive $D^{0}$-$\bar{D}^{\ast0}$ interaction induced by the mixing
has become too weak to create a bound state, the amplitude still has a pole close below
threshold, but on the unphysical Riemann sheet. This causes the big enhancement in the scaled cross-section close
above threshold, figure~\ref{scaledcsec100}.

When increasing $p$, the Argand diagram on the right panel in figure~\ref{argandD0D0asts} follows counterclockwise the unitary
circle. At an energy of about $3876$~MeV, the
modulus of $pf_{0}$ reaches a maximum with $\delta_{0}=\frac{\pi}{2}$. This
maximum corresponds to the peak of the first enhancement in the scaled cross-section, which reaches the value $1$. Then, the modulus of $pf_{0}$ decreases.
For an energy about 3880~MeV, corresponding to the opening of the $D^{+}D^{\ast-}$ channel, the Argand
diagram stops following the unitary circle, behaving in a completely analogous
manner to the bound state case analyzed above. This is also evident from the
$d$-wave Argand diagram, right panel in figure~\ref{argandD0D0astd}, showing the unaltered presence of the
Breit-Wigner resonance $\chi_{c1}(2p)$. Notice that, as in the bound state case, the $d$-wave contribution starts to play some significant role only after the $D^{+}D^{\ast-}$ channel has opened.

\begin{figure}
\centering
\includegraphics{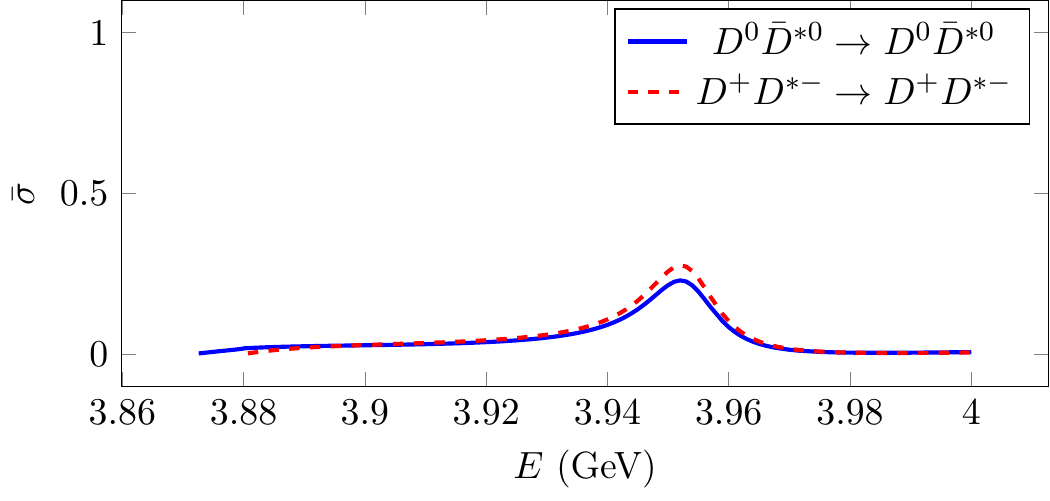}
\caption{\label{scaledcsec50}Elastic scaled cross-sections for $J^{PC}=1^{++}$, calculated with $\Delta_c=50$~MeV.}
\end{figure}

When further decreasing $\Delta_{c}$, the effect of the virtual state on the
Argand diagram and scaled cross-section becomes less relevant, so that for a
value $\Delta_{c}\leq50$~MeV it hardly plays a role, see figure~\ref{scaledcsec50}.
In contrast, the $\chi_{c1}(2p)$ enhancement is clearly visible as an isolated peak,
corresponding to a conventional $2p$ charmonium state.

\subsection{\texorpdfstring{$D^{+}D^{\ast-}\rightarrow D^{+}D^{\ast-}$}{D+D-starD+D-star} scattering}

\begin{figure}
\centering
\includegraphics{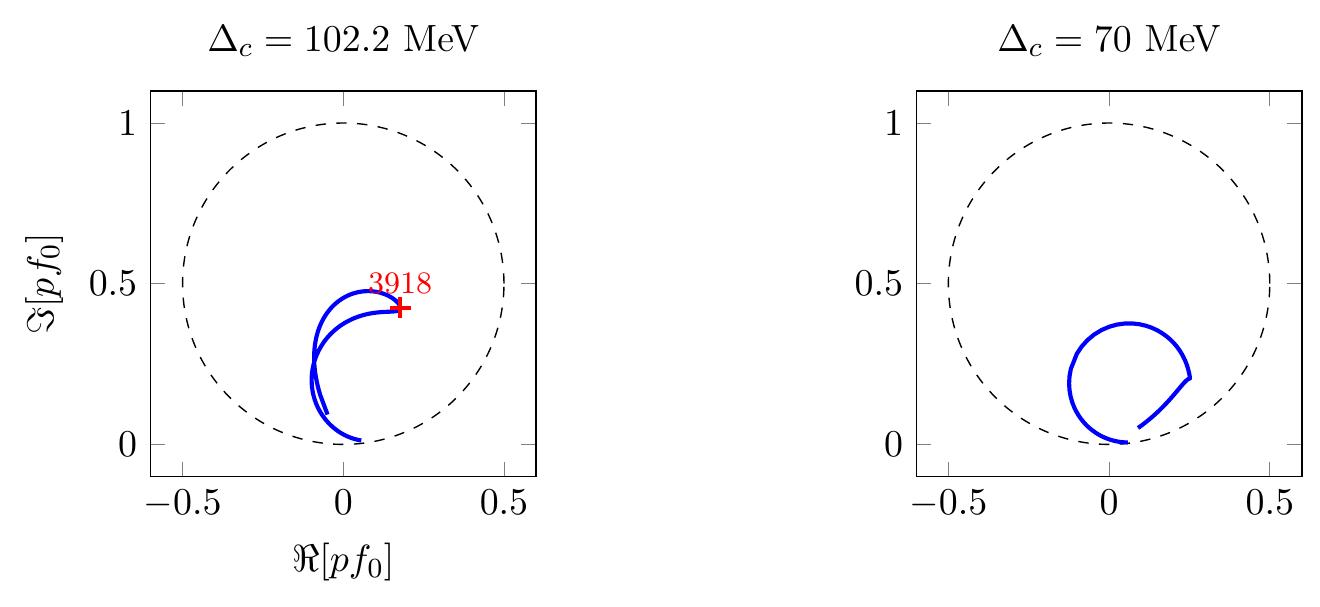}
\caption{\label{argandD+D-asts}Argand diagram for the $s$-wave elastic $D^+D^{\ast-}$ scattering amplitude, calculated with different values of $\Delta_c$. The energy at which the diagram with $\Delta_c=102.2$ MeV changes from clockwise to counterclockwise behavior, in MeV, is highlighted.}
\end{figure}

A look at figure~\ref{scaledcsec102} shows that the $D^{+}D^{\ast-}$ elastic,
scaled cross-section also presents two enhancements. However, the first
enhancement close above threshold is much less significant than that for the
elastic $D^{0}\bar{D}^{\ast0}$ scattering. On the one hand, this has to do
with the fact that the bound state assigned to $\chi_{c1}(3872)$ has a
$D^{+}D^{\ast-}$ component significantly smaller than the
$D^{0}\bar{D}^{\ast0}$ one, being also farther from the $D^{+}%
D^{\ast-}$ threshold than for the $D^{0}\bar{D}^{\ast0}$ one.
On the other hand, it is also related to the opening of the inelastic channel
$D^{+}D^{\ast-}\rightarrow D^{0}\bar{D}^{\ast0}$ at the
$D^{+}D^{\ast-}$ threshold. This makes, see left panel in figure~\ref{argandD+D-asts}, that the
$s$-wave Argand diagram at low momenta, showing a clockwise behavior, does not
follow the unitary circle, despite the negligible contribution from the $d$-wave.

\begin{figure}
\centering
\includegraphics{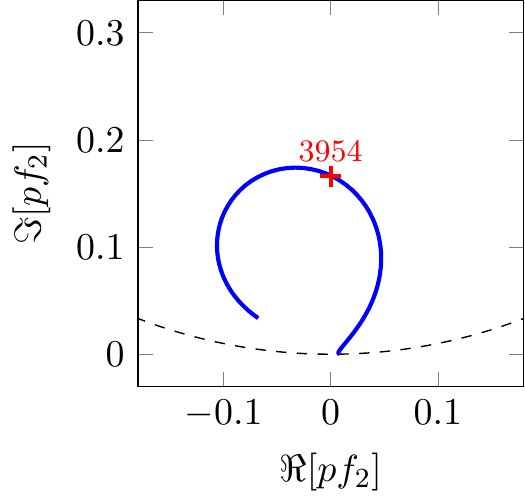}
\caption{\label{argandD+D-astd102}Argand diagram for the $d$-wave elastic $D^+D^{\ast-}$ scattering amplitude, calculated with $\Delta_c=102.2$~MeV. The energy at which $\delta_2=\pi/2$, in MeV, is highlighted.}
\end{figure}

For an energy about 3918~MeV,
the diagram changes to a counterclockwise behavior. This change is associated
to the presence of the resonance $\chi_{c1}(2p)$, as confirmed by the
$d$-wave Argand diagram, figure~\ref{argandD+D-astd102}.

This resonance, combined with the tail of the first enhancement, is
responsible for the second enhancement observed in the scaled cross-section
$\bar{\sigma}_{1^{++}}^{D^{+}D^{\ast-}\leftarrow
D^{+}D^{\ast-}}$, the slight increase of the value with respect to
$\bar{\sigma}_{1^{++}}^{D^{0}\bar{D}^{\ast0}\leftarrow
D^{0}\bar{D}^{\ast0}}$ indicating a slightly bigger coupling to the
$D^{+}D^{\ast-}$ channel than to the $D^{0}\bar{D}^{\ast0}$
one. This may be understood from the threshold mass difference between $D^{0}%
\bar{D}^{\ast0}$ and $D^{+}D^{\ast-}$. This difference makes
the $D^{+}D^{\ast-}$\nobreakdash-$c\bar{c}(2p)$ interaction
to be slightly more attractive than the $D^{0}\bar{D}^{\ast0}$\nobreakdash-$c\bar{c}(2p)$ one.

The threshold mass difference gives also rise, when changing $\Delta_{c}$, to
the appearance of a virtual state for a value of $\Delta_{c}$ below 80~MeV
instead of $\Delta_{c}\approx100$~MeV for $D^{0}\bar{D}^{\ast0}\rightarrow
D^{0}\bar{D}^{\ast0}$, as indicated by the counterclockwise behavior of
the $s$-wave Argand diagram at low momentum, right panel in figure~\ref{argandD+D-asts}. For higher energies, the
diagram confirms once more the persistence of the resonance $\chi_{c1}(2p)$.

When decreasing $\Delta_{c}$ even more, this is for $\Delta_{c}=50$~MeV and below,
the corresponding highly virtual state has almost no effect on the Argand diagram and
the scaled cross-section.

\section{Alternative scenarios}
\label{toysec}

In the previous analysis, we have assumed the $c\bar{c}$--meson-meson mixing induced by string breaking to be the very dominant interaction underlying the $D\bar{D}^\ast$ scattering process. However, the effective value of $\Delta_c$ could be implicitly taking into account, up to a certain extent, the effect of other neglected interactions. Next, we briefly analyze two alternative scenarios within the diabatic framework. For this purpose, our starting point in both case will be a string breaking mixing interaction whose strength parameter is taken as a flavor-independent constant fixed to the lattice value,
\[
\Delta_\textup{Lattice} \approx 51 / \sqrt{2} \text{ MeV} \approx 36.1\text{ MeV,}
\]
consistently with the understanding of the energy levels being independent of the heavy-quark mass aside a constant shift to all levels.

\subsection{Diabatic toy model with pion exchange}
\label{pitoy}


Let us consider the addition of a meson-meson interaction mediated by one pion exchange (OPE) represented by a simplified radial Yukawa-like potential,
\[
V_\textup{OPE}(r) = - g \frac{e^{- m_\pi r}}{r}
\]
so that the meson-meson diagonal diabatic potential matrix elements become
\[
T^{(i)} \to T^{(i)} + V_\textup{OPE}(r),
\]
with $g$ an effective coupling constant and $m_\pi$ the exchanged pion mass. As we are more interested in elastic scattering processes, we consider only exchanges of a neutral pion with a mass $m_{\pi}\approx135$ MeV, and we neglect charged pion exchanges mediating the inelastic $D^{0}\bar{D}^{\ast 0} \leftrightarrow D^+D^{\ast -}$ processes. As for the coupling constant $g$, we fix it requiring that altogether the mixing with $c\bar{c}$, with a strength fixed from lattice QCD, and the OPE potential generate the $\chi_{c1}(3872)$ close below the $D^0\bar{D}^{\ast 0}$ threshold. The resulting value is $g\approx 23.7$~MeV~fm, giving rise to a very dominant meson-meson probability.

Regarding this value, it may be taking into account the effect of terms that have not been considered in our simplified interaction. Actually, the value of $g$ is about four times bigger than the coupling constant $\frac{\gamma V_{0}}{m_{\pi}}\hbar c=5.7$ MeV fm used in \cite{Tor04} to describe $\chi_{c1}(3872)$ as a deuson with a binding energy of 500 keV from pion exchange assuming isospin invariance. Although a comparison is very difficult due to the quite different frameworks involved, the bigger value of $g$ could be somehow expected from the approximations followed (only $\pi^{0}$ exchange, no tensor interaction, no $D^{0}\bar{D}^{\ast0}$-$D^{+}D^{\ast-}$ diabatic potential mixing, etc\dots).

This effective OPE potential is by no means an accurate representation of the $D^{0}\bar{D}^{\ast 0}$ interaction (for a critical analysis of pion exchange between charm mesons, see, for instance, \cite{Tho08} and references therein). However, this toy model may be illustrative about the effects in more refined treatments.

\begin{figure}
\centering
\includegraphics{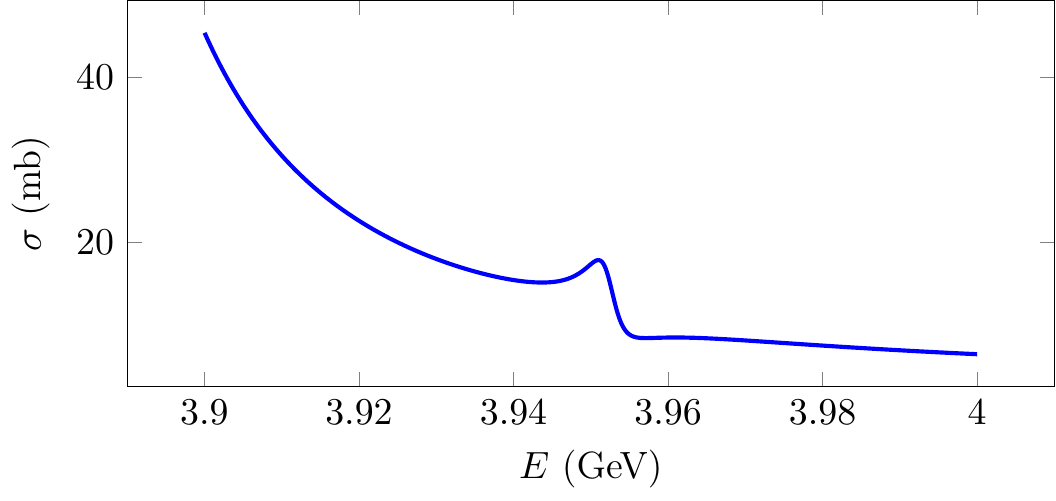}
\caption{\label{toycsec}Elastic $D^{0}\bar{D}^{\ast0}$ cross-section for $J^{PC}=1^{++}$ near 3960~MeV, calculated in the diabatic toy model with $\Delta_c = \Delta_\textup{Lattice}$ and simplified OPE potential.}
\end{figure}
In this toy model, the two separate scattering interactions, the one taking place through $c\bar{c}$--meson-meson mixing and that through $\pi_0$ exchange, add coherently in the scattering amplitude, from which the cross-sections are straightforwardly calculated.

Exactly like before, the elastic open-charm cross-sections for $J^{PC}=1^{++}$ calculated in the toy model show important enhancements at the $D^0 \bar{D}^{\ast 0}$ and $D^+D^{\ast -}$ thresholds. The $\chi_{c1}(2p)$ appears again as a small resonant structure in the elastic $D^0 \bar{D}^{\ast 0}$ cross-section near $3960$ MeV, see figure~\ref{toycsec}. The calculated resonance is narrower than the corresponding one in figure~\ref{natcsec102}, which is somehow expected since the mixing potential mediates the decay of $\chi_{c1}(2p)$ to open charm meson-meson channels and $\Delta_\textup{Lattice}$ is about one third of $\Delta_{c}$. This sharpness also allows to appreciate the typical behavior of the cross-section in presence of a resonance on a big background. This aside, the plot in figure~\ref{toycsec} shows no qualitative difference with respect to that in figure~\ref{natcsec102}, being the $\chi_{c1}(2p)$ structure completely dwarfed by the nearby threshold enhancements.

\subsection{Diabatic toy model with an additional compact state}

Thus far, we have shown that a molecular $\chi_{c1}(3872)$ may overshadow $\chi_{c1}(2p)$ in its open-charm di-meson decay channels. However, the nature of $\chi_{c1}(3872)$ is still matter of intense debate. On one hand, some of its decay properties, like the ratio of its  branching fractions to  $\rho J\!/\!\psi$ and $\omega J\!/\!\psi$, and the proximity of its mass to the $D^0\bar{D}^{\ast0}$ threshold, have been interpreted as hints towards a molecular nature, see, for instance, \cite{Hosaka16,Dong17,Guo18} and references therein. But other properties, like its observed prompt production in hadron collisions or a tentative determination of the sign of the $D^0\bar{D}^{\ast0}$ effective range, have been used as arguments in favor of a more compact quark nature, see, for instance, \cite{Big09,Esp22} and references therein.

Although the phenomenological nature of our diabatic potential prevents us from drawing a definite conclusion about the composition of $\chi_{c1}(3872)$, we can exploit the diabatic framework to contribute to such discussion, at least in what concerns the overshadowing of $\chi_{c1}(2p)$.

In principle, tetraquark channels and their potentials could be calculated on the lattice and then included in a diabatic scheme. While a  rigorous study is clearly beyond the reach of this paper, here we can at least qualitatively discuss the expected consequences on the overshadowing of $\chi_{c1}(2p)$ if one assumes a compact $\chi_{c1}(3872)$ nature in a diabatic toy model. Specifically, we proceed as in section~\ref{pitoy}, fixing the $c\bar{c}$-$D\bar{D}^\ast$ mixing strength parameter to the lattice value but  inserting an extra ``compact'' hidden-charm channel instead of adding a $D\bar{D}^\ast$ potential. We adjust the numerical values of the diabatic potential matrix elements involving the extra compact channel to have a pure compact state, with a mass about that of $\chi_{c1}(3872)$, whose mixing with $D\bar{D}^\ast$ generates a state with a mass close below the $D^0\bar{D}^{\ast0}$ threshold (at $3870.1$~MeV, to be precise) and a composition dominated by the additional compact channel (98\% of compact hidden-charm and 2\% of molecular $D^0\bar{D}^{\ast0}$ probabilities, to be precise).

\begin{figure}
\centering
\includegraphics{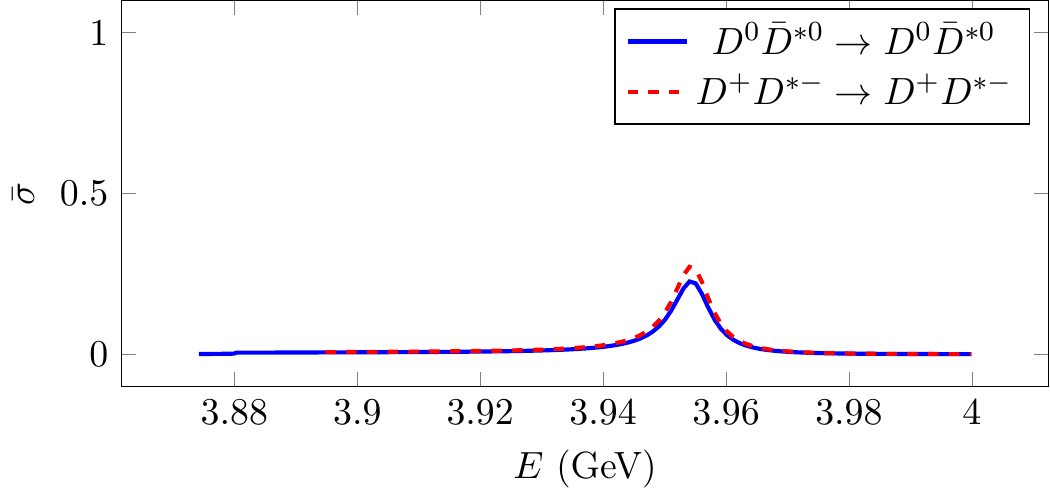}
\caption{\label{toycsec2}Elastic scaled $D^0\bar{D}^{\ast0}$ and $D^+D^{\ast-}$ cross-sections for $J^{PC}=1^{++}$, calculated in the diabatic toy model with $\Delta_c = \Delta_\textup{Lattice}$ and a compact $\chi_{c1}(3872)$.}
\end{figure}

The scaled elastic $D^0\bar{D}^{\ast0}$ and $D^+D^{\ast-}$ cross-sections calculated in this diabatic toy model are plotted in figure~\ref{toycsec2}. In this figure, the threshold enhancements are clearly absent and the $\chi_{c1}(2p)$ is immediately evident as an isolated peak. In this respect, the plot closely resembles that in figure~\ref{scaledcsec50}, with the only difference being a narrower calculated $\chi_{c1}(2p)$ peak, due to the fact that here we have used smaller value of $\Delta_c$ (36.1~MeV instead of 50~MeV). Therefore, we see that a compact $\chi_{c1}(3872)$ may not create enough threshold enhancement to overshadow $\chi_{c1}(2p)$, its effect being comparable to that of a highly virtual molecular state (see the end of section~\ref{pardep}).

\section{\texorpdfstring{$\chi_{c1}(2p)$}{chi c1 (2p)}}
\label{chisec}

The persistence of $\chi_{c1}(2p)$ even if no threshold enhancement were
present in the cross-section is explained because this resonance has a very
dominant $2p$ $c\bar{c}$ component coming from the interaction of the
$2p$ Cornell state with the continuum of $D^{0}\bar{D}^{\ast0}$ and
$D^{+}D^{\ast-}$ states through the diabatic mixing. In other
words, this resonance would be present (with a slightly different mass)
even if the mixing were not giving rise to $\chi_{c1}(3872)$. This seems to contradict the current experimental situation since no clear distinctive signal of $\chi_{c1}(2p)$ has been observed.

In this regard, we have shown that, assuming a molecular nature for $\chi_{c1}(3872)$, a pronounced cross-section enhancement close above the $D^{0}\bar{D}^{\ast0}$ and $D^{+}D^{\ast-}$ thresholds takes place, and this
prevents having a clear Breit-Wigner signal of $\chi
_{c1}(2p)$ in these channels. Instead, a very soft enhancement over a huge background, practically imperceptible in
the cross-section, makes difficult the experimental disentanglement of this
resonance through its expected dominant decays to $D^{0}\bar{D}^{\ast0}$ and
$D^{+}D^{\ast-}$. In this sense, accurate measurements of the
$D\bar{D}^{\ast}$ mass distribution around the energy of the resonance
would be very relevant. Notice though that this difficulty is specific for the $D^{0}\bar
{D}^{\ast0}$ and $D^{+}D^{\ast-}$ channels because they enter as
bound state components in $\chi_{c1}(3872)$. This suggests that other decay channels, for which overshadowing from $\chi_{c1}(3872)$
may not occur, might be more suited for the discovery of this resonance.

Regarding strong processes, it is natural to think of the OZI-forbidden decay
to $\omega J/\psi$. Indeed, for $\chi_{b1}(2p)$, which can be considered as the
bottomonium partner of $\chi_{c1}(2p)$, the decay to $\omega\Upsilon(1s)$ is significant (as it cannot decay to open-bottom channels). Moreover, although the decay $\chi_{c1}(3872) \to \omega J\!/\!\psi$ is significant, the presumably sharp lineshape of $\chi_{c1}(3872)$ makes overshadowing in this channel less likely. (Notice that this decay is expected to occur dominantly through the molecular components of $\chi_{c1}(3872)$, since $(c\bar{c}) \to \omega J\!/\!\psi$ is OZI-suppressed.) As a matter of fact, a recent study from BESIII
\cite{Abl19} does hint at a possible resonance with a
mass of $3963.7\pm5.5$~MeV and a quite uncertain width, $33.3\pm34.2$~MeV, for a
better fitting of the $\omega J/\psi$ mass distribution in $e^{+}%
e^{-}\rightarrow\gamma\omega J/\psi$.

As for electromagnetic processes involving $\chi_{c1}(2p)$, the decays
$\chi_{c1}(2p)\rightarrow\gamma\psi(2s)$ and $\chi_{c1}(2p)\rightarrow
\gamma J/\psi$ could have some significance, in parallel with the observed
decays of $\chi_{b1}(2p)$. On the other hand, we may expect a significant
production of $\chi_{c1}(2p)$ through $\psi(4040)\rightarrow\chi
_{c1}(2p)\gamma$ if $\psi(4040)$ contains, as leptonic width data seem to
indicate, a dominant $3s$ $c\bar{c}$ component. In this regard, current
data, giving only an upper bound ($3.4\times10^{-3}$) for the $\psi
(4040)\rightarrow\chi_{c1}\gamma$ branching fraction, are clearly
insufficient. We encourage an additional experimental effort along this line.


Additional indirect support to the existence of $\chi_{c1}(2p)$ could be obtained from
the discovery of expected $(0,2)^{++}$ charmoniumlike partners, this is, quasiconventional $c\bar{c}$ states with masses close to those of the $0^{++}$
and $2^{++}$, $2p$ $c\bar{c}$ states. However, at its current stage the
diabatic approach can neither accurately predict their masses nor their
possible overshadowing in the cross-sections by close structures, since it does not incorporate
spin-dependent corrections to the Cornell
potential (nor, consistently, other spin-dependent corrections in the diabatic
potential matrix). For
instance, the application of these corrections could make the $0^{++}$ $2p$
$c\bar{c}$ state to be below the $D_{s}\bar{D}_{s}$ threshold, what
could alter the results obtained in \cite{Bru21c}. As for
the $2^{++}$ $2p$ $c\bar{c}$ state, a similar situation may occur since
its mass could be increased to make it to be above the $D^{\ast0}\bar
{D}^{\ast0}$ threshold. The implementation of these corrections, out of the
scope of this article, will be the subject of a future study.

Therefore, we may conclude that there is a sound theoretical support and some
experimental indication in favor of the existence of $\chi_{c1}(2p)$, a
$1^{++}$ quasiconventional charmoniumlike resonance whose experimental
detection through its expected dominant $D^{0}\bar{D}^{\ast0}$ and $D^{+}%
D^{\ast-}$ decay modes may be hindered by the presence of the well-established
$\chi_{c1}(3872)$. More (and more accurate) data are needed to
confirm or refute this prescription.

With respect to this, the theoretical study carried out in Section~\ref{toysec} shows clearly that the overshadowing of $\chi_{c1}(2p)$ is a direct consequence of:
\begin{enumerate}[(i)]
\item\label{point1} the predicted mass of the $2p$ charmonium state being relatively close above the $D^0\bar{D}^{\ast 0}$ and $D^+D^{\ast-}$ thresholds;
\item\label{point2} string breaking coupling the $2p$ charmonium state with the continuum of $D^0\bar{D}^{\ast 0}$ and $D^+D^{\ast-}$ states;
\item\label{point3} the existence of a charmoniumlike meson state with dominant $D^0\bar{D}^{\ast 0}$ and $D^+D^{\ast-}$ components and a mass close below the corresponding thresholds.
\end{enumerate}

It should be pointed out that (\ref{point1}) is well-established from phenomenological potential models and underpinned by quenched lattice QCD calculations of the quarkonium potential, while (\ref{point2}) is supported by observation of string breaking in unquenched lattice QCD. As for (\ref{point3}), the $\chi_{c1}(3872)$ is experimentally well-established although its nature as a molecular or compact state is still under debate \cite{Hosaka16,Dong17,Guo18,Big09,Esp22,Baru22}. In this sense, the experimental observation of the overshadowing of $\chi_{c1}(2p)$ would constitute a compelling signal of the presence of a considerable molecular component in $\chi_{c1}(3872)$.

\section{Summary}
\label{sumsec}

We have carried out a thorough study of the $1^{++}$ $D^{0}\bar{D}%
^{\ast0}$ and $D^{+}D^{\ast-}$ elastic scattering, for center of
mass energies up to 4~GeV, in the so-called diabatic approach in QCD.

This formalism allows for the implementation of quenched and unquenched
lattice results for the energies of static quark and antiquark sources,
through the form of the diabatic potential matrix entering in a multichannel
Schr\"{o}dinger equation involving $c\bar{c}$ and meson-meson components.
Then, under the assumption that the meson-meson scattering takes place
predominantly through the mixing with $c\bar{c}$, namely meson-meson $\rightarrow
c\bar{c}\rightarrow$ meson-meson, the asymptotic solution of the
Schr\"{o}dinger equation provides us with the meson-meson scattering amplitude
and cross-section from the lattice dynamical input.

We have shown, through a detailed analysis of the scattering amplitudes under variation of the mixing strength parameter,
that independently of the character (bound or virtual state) of the
well-established $\chi_{c1}(3872)$, which has been assigned to a calculated
diabatic state, a quite robust prediction of a not yet established
resonance with a mass at about 3960~MeV comes out. The underlying physical
picture is that due to the vicinity of the $2p$ $c\bar{c}$ state to
the $D^{0}\bar{D}^{\ast0}$ and $D^{+}D^{\ast-}$ thresholds,
the meson-meson mixing with $c\bar{c}$, which is only relevant at
energies around the meson-meson thresholds, gives rise to two states, one at
an energy close to the threshold being predominantly of $D^{0}\bar
{D}^{\ast0}$ type, the $\chi_{c1}(3872)$, and the other at an energy close to that
of the $2p$ pure $c\bar{c}$ state being predominantly of $c\bar{c}$
type, the $\chi_{c1}(2p)$. Curiously, the vicinity of $\chi_{c1}(3872)$
prevents having a clearly distinctive signal of $\chi_{c1}(2p)$ in the
elastic $D^{0}\bar{D}^{\ast0}$ and $D^{+}D^{\ast-}$ cross-sections,
despite the fact that $\chi_{c1}(2p)$ is expected to decay very dominantly
through these channels. This difficulty may be overcome through the
consideration of alternative decay channels, for which the $\chi_{c1}(3872)$
might not overshadow the $\chi_{c1}(2p)$. This could be the case for the
$\omega J/\psi$ decay channel, according to a recent experimental exploration
of the $\omega J/\psi$ mass distribution from $e^{+}e^{-}\rightarrow
\gamma\omega J/\psi$ whose better fitting is achieved when a contribution from
a resonance with a mass at about 3960~MeV, which we tentatively identify with $\chi
_{c1}(2p)$, is taken into consideration.

As a possible alternative dynamical scenario, we have considered a diabatic toy model where a direct meson-meson interaction, explicitly neglected in the previous diabatic analysis, has been modeled through an effective one pion exchange potential. The cross-section calculated in this toy model shows that the overshadowing of $\chi_{c1}(2p)$ is not qualitatively sensitive to the details of the scattering potential but rather to the molecular composition of $\chi_{c1}(3872)$. This has been confirmed through the consideration of another diabatic toy model where $\chi_{c1}(3872)$ has been assumed to have a compact  nature (which could be simulating a tetraquark one) and no overshadowing for $\chi_{c1}(2p)$ is predicted.

Hence, the observation of the overshadowing of $\chi_{c1}(2p)$ would also provide a model-independent insight to the nature of $\chi_{c1}(3872)$.

\acknowledgments{
This work has been supported by \foreignlanguage{spanish}{Ministerio de Ciencia e Innovaci\'on} and \foreignlanguage{spanish}{Agencia Estatal de Investigaci\'on} of Spain MCIN/AEI/10.13039/501100011033 and European Regional Development Fund Grant No.~PID2019-105439~GB-C21, by EU Horizon 2020 Grant No.~824093 (STRONG-2020), and by \foreignlanguage{spanish}{Conselleria de Innovaci\'on, Universidades, Ciencia y Sociedad Digital, Generalitat Valenciana} GVA~PROMETEO/2021/083. R.B. acknowledges a \foreignlanguage{spanish}{Formaci\'on de Personal Investigador} fellowship from \foreignlanguage{spanish}{Ministerio de Ciencia, Innovaci\'on y Universidades} of Spain under Grant No.~BES-2017-079860. We thank Eric Braaten for his constructive comments.
}

\bibliography{overshadowbib}

\end{document}